\documentclass[conference]{IEEEtran}
\IEEEoverridecommandlockouts
\usepackage{cite}
\usepackage{amsmath,amssymb,amsfonts}
\usepackage{algorithmic}
\usepackage{graphicx}
\usepackage{textcomp}
\usepackage{xcolor}
\usepackage[colorlinks=false]{hyperref}
\usepackage{booktabs}
\usepackage{multirow}
\usepackage{float}
\usepackage{array}
\usepackage{listings}
\usepackage{tabularx}
\usepackage[bottom]{footmisc}

\def\BibTeX{{\rm B\kern-.05em{\sc i\kern-.025em b}\kern-.08em
    T\kern-.1667em\lower.7ex\hbox{E}\kern-.125emX}}

\begin{document}

\title{Adversarially Robust and Interpretable Magecart Malware Detection}

\author{
\IEEEauthorblockN{Pedro Pereira\textsuperscript{*}}
\IEEEauthorblockA{\textit{GECAD, ISEP, Polytechnic of Porto}\\
Porto, Portugal \\
0009-0008-7641-1566}
\and
\IEEEauthorblockN{José Gouveia}
\IEEEauthorblockA{\textit{GECAD, ISEP, Polytechnic of Porto}\\
Porto, Portugal \\
0009-0000-3673-4450}
\and
\IEEEauthorblockN{João Vitorino}
\IEEEauthorblockA{\textit{GECAD, ISEP, Polytechnic of Porto}\\
Porto, Portugal \\
0000-0002-4968-3653}
\and
\IEEEauthorblockN{Eva Maia}
\IEEEauthorblockA{\textit{GECAD, ISEP, Polytechnic of Porto}\\
Porto, Portugal \\
0000-0002-8075-531X}
\and
\IEEEauthorblockN{Isabel Praça}
\IEEEauthorblockA{\textit{GECAD, ISEP, Polytechnic of Porto}\\
Porto, Portugal \\
0000-0002-2519-9859}
\thanks{\textsuperscript{*}Corresponding author: peesp@isep.ipp.pt}
}

\maketitle

\begin{abstract}
Magecart skimming attacks have emerged as a significant threat to client-side security and user trust in online payment systems. This paper addresses the challenge of achieving robust and explainable detection of Magecart attacks through a comparative study of Machine Learning (ML) models with a real-world dataset. Tree-based, linear, and kernel-based models were applied with hyperparameter tuning and feature selection, to distinguish between benign and malicious scripts. The models are supported by a Behavior Deterministic Finite Automaton (DFA), which captures structural behavior patterns in scripts, helping to analyze and classify client-side script execution logs. To ensure robustness against adversarial evasion attacks, adversarial training and evaluations were performed using attacks from Adversarial Robustness Toolbox and Adaptative Perturbation Pattern Method. In addition, concise explanations of ML decisions are provided, supporting transparency and user trust. Experimental validation demonstrated high detection performance and interpretable reasoning, demonstrating that traditional ML models can be effective in real-world web security contexts.
\end{abstract}


\section{Introduction}
In recent years, the complexity and frequency of cyberattacks targeting online payment systems have risen sharply~\cite{attacks}. Among them, Magecart skimming attacks have become a major threat, exploiting vulnerabilities to inject or modify malicious scripts on e-commerce platforms~\cite{toreini2019}. Between 2022 and 2023, United States of America authorities reported a 96\% increase in e-Commerce skimming incidents~\cite{Cobb_2023}. The emergence of ``Magecart-as-a-Service'' has further simplified attacks, for instance, the ``Sniffer by Fleras'' kit, sold for \$1,500 on dark web forums, compromised over 480 websites in early 2024~\cite{Fiesel_2024}. These attacks remain difficult to detect, posing serious risks to client-side security and user trust~\cite{client}.

This paper explores the use of Machine Learning (ML) classification models combined with a Behavior Deterministic Finite Automaton (DFA)~\cite{pereira2024dfa} to enhance Magecart detection. The Behavior DFA models JavaScript execution structures, enriching ML training with behavioral features. Model training and optimization were conducted on proprietary client-side JavaScript execution logs from multiple e-Commerce platforms using hyperparameter tuning and feature selection.

Model robustness against evasion was assessed with adversarial attacks from the Adversarial Robustness Toolbox (ART)~\cite{nicolae2019adversarialrobustnesstoolboxv100_ART} and the Adaptative Perturbation Pattern Method (A2PM)~\cite{vitorino2022adaptative_a2pm}. Realistic adversarial conditions were simulated by introducing subtle data perturbations with black-box and white-box adversarial evasion attacks, as well as with A2PM for targeted model-specific perturbations. The attacks included HopSkipJump, Boundary Attack, Fast Gradient Sign Method (FGSM) and Projected Gradient Descent (PGD).

Beyond achieving robust detection, this study also focused on providing interpretable explanations for the models' predictions. SHapley Additive exPlanations (SHAP), a statistical method to show how features influence predictions, was combined with a symbolic analysis with the Behavior DFA, modeling script execution paths and behavioral patterns. The insights from both methods were integrated and translated by a Large Language Model (LLM) into simple, natural language summaries. This enables the explanation of which actions occurred and how they led to a certain decision.

This study makes three key contributions: (1) enhanced Magecart detection by combining ML and behavioral DFA features; (2) improved robustness against evasion through adversarial training and evaluation; and (3) interpretable, explainable model outputs. Together, these establish reliable and transparent detection of malicious client-side scripts in real-world e-Commerce environments.

\hyperref[sec:2]{Section 2} reviews related work on Magecart detection, explainability, and robustness. \hyperref[sec:3]{Section 3} details the methodology, \hyperref[sec:4]{Section 4} presents results, and \hyperref[sec:5]{Section 5} concludes with key insights and future work.

\section{Related Work}\label{sec:2}
Current detection methods against Magecart attacks rely mainly on static and dynamic analysis, each with inherent limitations~\cite{zozzle,Prophiler,jsand}.
Static approaches examine downloaded JavaScript to identify known signatures or suspicious patterns through syntactic and lexical analysis. Tools such as ZOZZLE apply Bayesian classification on abstract syntax tree features to detect malware~\cite{zozzle}, while Prophiler leverages ML on HTML, JavaScript, and URL features to filter benign web pages~\cite{Prophiler}. Although static analysis is fast and scalable, it struggles to detect obfuscated or dynamically generated scripts, which do not reveal obvious static patterns. In contrast, dynamic analysis executes web pages in sandboxed environments to monitor runtime behavior, identifying activities such as form data exfiltration. Tools like JSAND~\cite{jsand} detect malicious actions by observing script execution and data flow. This approach can reveal previously unknown threats but is resource-intensive and vulnerable when malware detects the analysis environment.

Beyond code analysis, behavioral detection methods analyze how scripts interact with the browser and network. Recent research introduced automata-based models to represent behavioral patterns. Building on traditional Finite-State Automata (FSA), Pereira et al.~\cite{pereira2024dfa} proposed a weighted Behavior DFA to model JavaScript execution sequences and assign risk-based weights to actions. This method produces interpretable outputs and demonstrated effective real-world detection, combining structure and transparency in malware identification.

The integration of Machine Learning (ML) has also significantly improved JavaScript malware detection~\cite{review}. Studies have shown that ML models enhance precision in identifying malicious scripts. For instance, ensemble and kernel-based classifiers have achieved high detection rates in identifying vulnerable or injected code. Deep learning models, including dense neural networks and LSTMs, have been applied to detect Magecart-like behaviors by analyzing event sequences such as script injection, DOM manipulation, and data exfiltration. However, ML models often operate as black boxes, limiting analyst understanding of their classification rationale~\cite{ribeiro2016}. 

To address this, explainable AI (XAI) techniques have been introduced to make model predictions more interpretable. SHAP is a leading XAI method that quantifies the precise contribution of each input feature to a model's output. By assigning an importance value to each feature for every individual prediction, SHAP allows analysts to understand exactly how a model arrived at its decision. For instance Younas et al.~\cite{YOUNAS2024100466} used SHAP to interpret a Random Forest (RF) model for XSS detection, enhancing the transparency and reliability of automated analysis. Specifically, the authors used a SHAP chart to visualize how each feature contributed to the model's predictions. The chart displayed each feature's SHAP value, showing whether it pushed the prediction toward an attack or benign classification compared to a baseline, allowing analysts to identify the most influential patterns in XSS detection.

At the same time, robustness is critical against adversarial manipulation. Attackers increasingly use obfuscation, code fragmentation, and loader scripts to evade detection. Systems like JSRevealer~\cite{10202610} have shown high accuracy against heavily obfuscated JavaScript. Beyond such specialized systems, robustness can be further enhanced through adversarial training, ensemble learning, and runtime feature analysis~\cite{chen2022}

In summary, existing research demonstrates that while static and dynamic analyses provide foundational techniques, they struggle with modern, evasive attacks. Combining ML with behavioral modeling offers more effective detection, but future Magecart defense systems must also ensure robustness and explainability to achieve reliable and interpretable threat detection in adversarial web environments.

\section{Methods}\label{sec:3}

\subsection{Data Preprocessing} 

This study analyzes a proprietary dataset capturing the execution of JavaScript scripts from real-world e-commerce platforms. The dataset logs script operations across the shopping journey, from login and browsing to checkout, providing fine-grained, session-level records of actions such as modifying HTML attributes, making network requests, or handling user interactions. Contextual information, session, page, and temporal order, is also included. From these actions, behaviors were derived to represent their actual effects on the platform. While an action describes what a script performs, a behavior captures its impact, such as DOM manipulation, event handling, or data transmission. As scripts execute multiple actions, they can exhibit different behaviors throughout their lifecycle. This study is focused on behaviors, as they directly reflect the effects of script execution on the shopping journey. For each script, a sequence of behaviors was extracted, preserving their temporal order. These behavioral sequences reveal whether a script acts benignly or maliciously.

Two groups of features were engineered to capture both behavioral dynamics and contextual factors. Behavioral features include: the sequence length, representing the total number of behaviors; positional indicators (first and last normalized positions), which describe where each behavior occurs within the sequence (0.0 = start, 1.0 = end, -1.0 = absent); and behavior prevalence, quantifying how frequently each behavior appears and thus reflecting its relative importance in the script’s logic. Contextual features capture external conditions influencing script execution, namely the average number of user clicks preceding script activation and a binary flag indicating whether the script was dynamically injected or originally embedded in the page.

Finally, following~\cite{pereira2024dfa}, a weighted Behavior DFA was used to analyze execution patterns and assign risk-based weights to behaviors for detecting malicious sequences. Unlike the original DFA, which required the first element to match the initial state, the improved version allows any subsequence to be analyzed, increasing detection capacity by identifying malicious patterns at varying positions within the sequence. The resulting outputs, combined with the extracted features, produced a final dataset of 103 features, which was then used to train the ML models.

\subsection{Machine Learning Models}

All experiments were executed on an AMD Ryzen 5 5600 CPU and 32 GB RAM. Performing all tasks on a single machine ensured consistency and reproducibility throughout model training and evaluation. The following ML models were selected: Decision Tree (DT)~\cite{popat2018survey_DT_SVM_LR}, RF~\cite{rf_for_attack_detection}, Ada Boost Classifier (AdaBoost)~\cite{Nagarjun2020}, Gradient Boosting Classifier (GB)~\cite{Nagarjun2020}, Logistic Regression (LR)~\cite{popat2018survey_DT_SVM_LR}, Support Vector Machine (SVM)~\cite{kazemian2015comparisons_SVM_KNN_KMEANS}, Gaussian Naive Bayes (NB)~\cite{wang2015jsdc_RF_NB}, K-Nearest Neighbors (KNN)~\cite{kazemian2015comparisons_SVM_KNN_KMEANS}, and K-means~\cite{kazemian2015comparisons_SVM_KNN_KMEANS}. These models were fitted with the best hyperparameters using a 5-fold cross-validation grid search, using F1 as the evaluation metric to guide the optimization process. This enables systematically evaluating a variety of hyperparameter combinations and selecting the optimal configuration for each model avoiding overfitting.

To train the ML models a feature selection process based on feature importance was employed. The importance of each feature was determined using a RF which was trained on the complete dataset, among the 103 features the 60 features that exhibited the highest importance scores were selected. The choice of 60 features was not arbitrary, it was selected from a series of experimental evaluations, in which subsets of features were tested ranging from 20 to 100, in increments of 10. This evaluation was carried out on all implemented ML models, analyzing the F1 metric. The selection of 60 features corresponded to the setting that produced the best overall F1.

\subsection{Adversarial Robustness Evaluation}

To evaluate the robustness and generalization of the ML models, adversarial attacks from the ART and A2PM methods were used. Specifically we used Hop Skip Jump~\cite{chen2020hopskipjumpattack_HSJ}, Boundary Attack~\cite{brendel2017decision_Boundary_attack}, FGSM~\cite{liu2019sensitivity_FGSM}, PGD~\cite{madry2017towards_PGD} and A2PM~\cite{vitorino2022adaptative_a2pm}. During training, each attack was used to generate new samples from the training dataset with minimal perturbations. These adversarial samples were combined with the original training dataset to help the models learn from manipulated inputs and improve their resilience. For evaluation, each attack was applied to the testing dataset to simulate real-world evasion attempts. The models were then tested separately on each adversarial set to assess their performance under different threat scenarios. This strategy enabled a granular analysis of the models' performance under varying adversarial scenarios, highlighting their respective strengths and weaknesses in a controlled and replicable setting.

\subsection{Interpretable Model Explanations}

To enhance the explainability of ML models, a hybrid approach is proposed, combining feature attribution, symbolic reasoning, and natural language generation. The first component uses SHAP~\cite{lundberg2017}, a model-agnostic method that quantifies each feature’s contribution to a prediction. For every classified script, SHAP values identify the most influential features, indicating whether they increased or decreased the likelihood of being malicious. Features are ranked by absolute contribution, producing structured summaries of the key decision factors (see Appendix~\ref{app:shap_output}).

Concurrently, the Behavior DFA assesses each script’s structural behavior by comparing observed action sequences to known malicious patterns. It outputs a match percentage, a classification label (Benign, Partially Malign, or Malign), and a breakdown of behavioral transitions with risk weights (see Appendix~\ref{app:dfa_output}).
This output highlights how the Behavior DFA not only recognizes the presence of high-risk behaviors, such as \textit{``Send Data''}, but also quantifies their cumulative impact based on a predefined risks. Such transparent and symbolic reasoning complements the statistical inferences of SHAP, offering a second layer of interpretability.

To synthesize both perspectives into coherent explanations, an instruction-tuned LLM, LLaMA-4 Scout~\cite{LLAMA4}, is employed. This LLM is prompted with structured input that includes both SHAP insights and Behavior DFA analysis.

\section{Results and Discussion}\label{sec:4}
To evaluate the performance of the ML models, two main types of metrics were used, quality metrics - Accuracy, Precision, Recall and F1 - and footprint metrics - training and prediction time, both in seconds.

The results presented in Table \ref{tab:resultados_modelos_fs_importancia} demonstrate the performance of the implemented ML models. To evaluate these results, F1 and Recall are identified as the most critical metrics, primarily due to the necessity of minimizing false negatives, malign entries classified as benign, within a security framework. From the results, it is possible to conclude that the SVM achieves a strong balance within quality metrics, with a high Recall of 0.9528, F1 of 0.9571, and overall Accuracy of 0.9964, making it the most reliable model to detect Magecart attacks. KNN also performs well with a Recall of 0.9428 and the fastest training time, 0.0010 s, although its Recall is slightly lower than SVM. LR and RF offer competitive results, with F1 above 0.90, however, they have slightly lower Recall values, 0.9041 and 0.8805, respectively, which could result in more missed detections. NB, while achieving a very high Recall, 0.9905, has extremely low Precision 0.1325, making it impractical due to an overwhelming number of false positives. In general, SVM emerges as the best option when using all features, providing excellent detection capability at a reasonable computational cost.

\begin{table}[h]
\centering
\caption{Obtained results with selected features}
\label{tab:resultados_modelos_fs_importancia}
\scriptsize
\setlength{\tabcolsep}{5pt}
\begin{tabular}{@{}cccccccc@{}}
\toprule
\textbf{Model} & \textbf{Acc.} & \textbf{Prec.} & \textbf{Rec.} & \textbf{F1} & \textbf{Train. Time} & \textbf{Pred. Time} \\ \midrule
DT & 0.9916 & 0.9337 & 0.8616 & 0.8948 & 0.0333 & 0.0005 \\
RF & 0.9936 & 0.9640 & 0.8805 & 0.9200 & 0.0818 & 0.0154 \\
AdaBoost & 0.9900 & \textbf{0.9788} & 0.7798 & 0.8654 & 1.2404 & 0.0430 \\
GB & 0.9918 & 0.9572 & 0.8423 & 0.8952 & 1.0358 & 0.0010 \\
LR & 0.9930 & 0.9268 & 0.9041 & 0.9148 & 0.0121 & 0.0006 \\
SVM & \textbf{0.9964} & 0.9528 & 0.9618 & \textbf{0.9571} & 0.0267 & 0.0102 \\
NB & 0.7274 & 0.1325 & \textbf{0.9905} & 0.2337 & 0.0024 & \textbf{0.0000} \\
KNN & 0.9950 & 0.9428 & 0.9380 & 0.9401 & \textbf{0.0010} & 0.0084 \\
K-means & 0.4600 & 0.0474 & 0.5271 & 0.0865 & 0.0259 & 0.0012 \\
\bottomrule
\end{tabular}
\end{table}

To assess the resilience of the models under adversarial conditions, Table \ref{tab:attack_results_by_model} presents the quality metrics after subjecting each model to various adversarial attacks. Bold numbers indicate the best-performing value for each metric within a model across the evaluated attacks. The results confirm a clear stratification of the robustness between the models.

\begin{table}[h]
\scriptsize
\setlength{\tabcolsep}{5pt}
\centering
\caption{Obtained results after adversarial attacks per model}
\label{tab:attack_results_by_model}
\begin{tabular}{llcccc}
\toprule
\textbf{Model} & \textbf{Attack} & \textbf{Acc.} & \textbf{Prec.} & \textbf{Rec.} & \textbf{F1} \\
\midrule

\multirow{3}{*}{\textbf{DT}} 
& HopSkipJump & \textbf{0.9532} & 0.0000 & 0.0000 & 0.0000 \\
& Boundary Attack & \textbf{0.9532} & 0.0000 & 0.0000 & 0.0000 \\
& A2PM & \textbf{0.9532} & 0.0000 & 0.0000 & 0.0000 \\
\midrule

\multirow{3}{*}{\textbf{RF}} 
& HopSkipJump & 0.9626 & 0.8182 & 0.1429 & 0.2432 \\
& Boundary Attack & 0.9599 & 0.7143 & 0.0794 & 0.1429 \\
& A2PM & \textbf{0.9686} & \textbf{0.9000} & \textbf{0.2857} & \textbf{0.4337} \\
\midrule

\multirow{3}{*}{\textbf{AdaBoost}} 
& HopSkipJump & \textbf{0.9525} & 0.0000 & 0.0000 & 0.0000 \\
& Boundary Attack & \textbf{0.9525} & 0.0000 & 0.0000 & 0.0000 \\
& A2PM & \textbf{0.9525} & 0.0000 & 0.0000 & 0.0000 \\
\midrule

\multirow{3}{*}{\textbf{GB}} 
& HopSkipJump & 0.9552 & 0.1667 & 0.0159 & 0.0290 \\
& Boundary Attack & \textbf{0.9559} & \textbf{0.2857} & \textbf{0.0317} & \textbf{0.0571} \\
& A2PM & \textbf{0.9559} & \textbf{0.2857} & \textbf{0.0317} & \textbf{0.0571} \\
\midrule

\multirow{5}{*}{\textbf{LR}} 
& HopSkipJump & \textbf{0.9947} & \textbf{0.9365} & \textbf{0.9365} & \textbf{0.9365} \\
& Boundary Attack & 0.9933 & 0.9344 & 0.9048 & 0.9194 \\
& FGSM & 0.9853 & 0.9184 & 0.7143 & 0.8036 \\
& PGD & 0.9853 & 0.9184 & 0.7143 & 0.8036 \\
& A2PM & 0.9612 & 0.6923 & 0.1429 & 0.2368 \\
\midrule

\multirow{5}{*}{\textbf{SVM}} 
& HopSkipJump & \textbf{0.9967} & \textbf{0.9677} & \textbf{0.9524} & \textbf{0.9600} \\
& Boundary Attack & 0.9960 & 0.9672 & 0.9365 & 0.9516 \\
& FGSM & 0.9926 & 0.9643 & 0.8571 & 0.9076 \\
& PGD & 0.9920 & 0.9636 & 0.8413 & 0.8983 \\
& A2PM & 0.9592 & 0.6667 & 0.0635 & 0.1159 \\
\midrule

\multirow{3}{*}{\textbf{Naive Bayes}} 
& HopSkipJump & \textbf{0.7380} & \textbf{0.1353} & \textbf{0.9683} & \textbf{0.2374} \\
& Boundary Attack & \textbf{0.7380} & \textbf{0.1353} & \textbf{0.9683} & \textbf{0.2374} \\
& A2PM & 0.6972 & 0.0000 & 0.0000 & 0.0000 \\
\midrule

\multirow{3}{*}{\textbf{KNN}} 
& HopSkipJump & 0.9612 & 0.7273 & 0.1270 & 0.2162 \\
& Boundary Attack & \textbf{0.9693} & \textbf{0.8696} & \textbf{0.3175} & \textbf{0.4651} \\
& A2PM & \textbf{0.9693} & \textbf{0.8696} & \textbf{0.3175} & \textbf{0.4651} \\
\midrule

\multirow{1}{*}{\textbf{K-means}} 
& A2PM & 0.3416 & 0.0011 & 0.0159 & 0.0020 \\
\bottomrule
\end{tabular}
\end{table}

DT and AdaBoost, exhibit major vulnerability, showing 0.0000 Precision, Recall, and F1 under all attacks despite Accuracy near 0.95, indicating complete failure to detect true positives under adversarial conditions. RF performs slightly better, reaching an F1 of 0.4337 in A2PM but still suffers from low Recall (0.2857). GB also shows limited ability to handle adversarial examples, with F1 below 0.06. LR behaves inconsistently, with low F1 and Recall under A2PM (0.2368 and 0.1429) but strong performance against other attacks, maintaining Recall above 0.70 and F1 above 0.80, showing resistance to gradient-based attacks. SVM stands out as the most robust model, maintaining F1 above 0.89 for all attacks except A2PM (0.1159) and achieving Recall above 0.84 and Precision above 0.96, confirming its suitability for adversarial environments. NB and KNN behave differently: NB attains high Recall (0.9683) but very low Precision (0.1353), while KNN achieves a better trade-off with F1 around 0.4651 in A2PM and Boundary Attack.

Overall, SVM remains the most resilient model, while DT, AdaBoost, GB, and K-means consistently underperform. A2PM causes the most severe degradation across all models, suggesting it generates more effective perturbations, while HopSkipJump and PGD show inconsistent effects. These results confirm that no single attack impacts all classifiers equally and highlight the need to evaluate detection models under multiple adversarial scenarios to accurately assess robustness and ensure reliability in real-world deployment.

To evaluate the effectiveness of the explainability methods, tests were conducted on scripts exhibiting behaviors from benign to potentially malicious. In one representative case, the script executed a sequence including setting callbacks, adding event handlers, accessing input and DOM attributes, creating DOM elements, and sending data. While such behaviors are common in legitimate contexts, specific patterns can indicate malicious intent.

In this case, the Behavior DFA classified the script as ``MALIGN'' with a 100\% match to known patterns, and the ML model assigned a 99.66\% malicious probability. SHAP analysis showed that behaviors like setting callbacks, adding handlers, and sending data most influenced the prediction. The generated explanation (Appendix \ref{app:llm-explanations-2}) summarized the findings in accessible language. This hybrid system effectively bridged complex model outputs and human understanding, enhancing transparency in security-critical contexts.

\section{Conclusions}\label{sec:5}

This study demonstrated that a robust and interpretable system for Magecart malware detection can be achieved by integrating ML models with an automaton-based classification framework. Modeling behavioral patterns through the Behavior DFA and combining them with ML models enables the system to capture both structural and statistical irregularities in client-side script executions.

The implemented ML models consistently demonstrated strong robustness against adversarial manipulation, ensuring reliable threat classification even under a diverse set of evasion attacks. Among the evaluated models, the SVM consistently achieved the highest results across performance metrics, especially on Recall. This indicates that it is suitable for contexts where false negatives pose significant risks. In addition to robustness, the importance of explainability is also emphasized. Explainable outputs were generated as natural language descriptions by an LLM, translating SHAP values and Behavior DFA insights into transparent justifications for the ML model's decisions.  Overall, the results indicated that the robust explainable models behaved as expected, being suitable for security-critical applications.

Future work will extend detection beyond Magecart to other JavaScript-based threats, including clickjacking and DOM-based XSS. Additional ML and DL models (e.g., LSTM, CNN) and an optimized, lightweight browser extension or edge-based deployment will be explored to enhance real-time performance and maintain explainability and robustness.

\section*{Acknowledgments}
This work was done and funded in the scope of the BEHAVIOR project (NORTE2030-FEDER-00576300 no. 14391). This work was also supported by UIDB/00760/2020.

\bibliographystyle{IEEEtran}
\bibliography{refs.bib}

\appendix

\section{Example Explanations and Outputs}

\subsection{SHAP Feature Contributions Example} \label{app:shap_output}
\begin{figure}[H]
\centering
\begin{lstlisting}[basicstyle=\ttfamily\footnotesize,breaklines=true]
Base prediction (expected value): -4.0197
Feature contributions:
- Automaton Classification: decreased the prediction by 0.8394 (value: 1.0)
- inline: decreased the prediction by 0.6822 (value: False)
...
- First_Replace DOM Element(s): decreased the prediction by 0.0000 (value: -1.0)
- Last_Inject Code Dynamically: decreased the prediction by 0.0000 (value: -1.0)
Final prediction: 0.9966 (probability of being malicious)
\end{lstlisting}
\end{figure}

\subsection{Behavior DFA Output Example} \label{app:dfa_output}
\begin{figure}[h]
\centering
\begin{lstlisting}[basicstyle=\ttfamily\footnotesize,breaklines=true]
-> Classification: MALIGN (100.00%)
-> Nearest Final State: q1719
-> Matched Symbols and Weights:
   - Set Callback (Weight: 3)
   - Add Event Handler (Weight: 3)
   - Access Input, Access DOM Element Attribute (Weight: 4)
   - Create DOM Element (Weight: 1)
   - Send Data, Update DOM Element (Weight: 5)
-> Total Matched Weight: 16
-> Match Percentage Calculation: 
(16 / 16) * 100 = 100.00%
\end{lstlisting}
\end{figure}

\section{Example LLM-Generated Explanations}

\subsection{Narrative Explanation} \label{app:llm-explanations-2}
\small\texttt{}
\begin{quote}
This script is classified as malicious with a high risk of approximately 99.66\%.

The Automata Model has identified a strong match to known malicious patterns, labeling it as MALIGN with a 100\% match percentage, primarily due to behaviors such as setting callbacks, adding event handlers, accessing input and DOM elements, creating DOM elements, and sending data, which are commonly seen in malicious scripts. 

The ML Model, which assesses risk based on specific features, also indicates a high risk, with features like adding DOM elements, sending data, and updating DOM elements contributing to the malicious classification, while safe behaviors like accessing known content or using standard features are not prominent; notably, there are no significant indications of benign activities.

Overall, the script appears to be harmful, with a high likelihood of performing malicious actions, and human review is recommended to understand its exact capabilities and mitigate potential threats.
\end{quote}

\end{document}